\documentclass[nofootinbib,aps,prd,preprint]{revtex4}
\usepackage{amssymb,amsmath,latexsym}

\usepackage{chemarr}
\usepackage{enumitem}
\usepackage{graphicx}
\usepackage{amssymb}
\usepackage{lineno}
\usepackage{amsmath}
\begin{document}
\date{\today}
\newcommand{\MSun}{{M_\odot}}
\newcommand{\LSun}{{L_\odot}}
\newcommand{\Rstar}{{R_\star}}
\newcommand{\calE}{{\cal{E}}}
\newcommand{\calM}{{\cal{M}}}
\newcommand{\calV}{{\cal{V}}}
\newcommand{\calO}{{\cal{O}}}
\newcommand{\calH}{{\cal{H}}}
\newcommand{\calD}{{\cal{D}}}
\newcommand{\calB}{{\cal{B}}}
\newcommand{\calK}{{\cal{K}}}
\newcommand{\labeln}[1]{\label{#1}}
\newcommand{\Lsolar}{L$_{\odot}$}
\newcommand{\farcmin}{\hbox{$.\mkern-4mu^\prime$}}
\newcommand{\farcsec}{\hbox{$.\!\!^{\prime\prime}$}}
\newcommand{\kms}{\rm km\,s^{-1}}
\newcommand{\cc}{\rm cm^{-3}}
\newcommand{\Alfven}{$\rm Alfv\acute{e}n$}
\newcommand{\Vap}{V^\mathrm{P}_\mathrm{A}}
\newcommand{\Vat}{V^\mathrm{T}_\mathrm{A}}
\newcommand{\D}{\partial}
\newcommand{\DD}{\frac}
\newcommand{\TAW}{\tiny{\rm TAW}}
\newcommand{\mm }{\mathrm}
\newcommand{\Bp }{B_\mathrm{p}}
\newcommand{\Bpr }{B_\mathrm{r}}
\newcommand{\Bpz }{B_\mathrm{\theta}}
\newcommand{\Bt }{B_\mathrm{T}}
\newcommand{\Vp }{V_\mathrm{p}}
\newcommand{\Vpr }{V_\mathrm{r}}
\newcommand{\Vpz }{V_\mathrm{\theta}}
\newcommand{\Vt }{V_\mathrm{\varphi}}
\newcommand{\Ti }{T_\mathrm{i}}
\newcommand{\Te }{T_\mathrm{e}}
\newcommand{\rtr }{r_\mathrm{tr}}
\newcommand{\rbl }{r_\mathrm{BL}}
\newcommand{\rtrun }{r_\mathrm{trun}}
\newcommand{\thet }{\theta}
\newcommand{\thetd }{\theta_\mathrm{d}}
\newcommand{\thd }{\theta_d}
\newcommand{\thw }{\theta_W}
\newcommand{\beq}{\begin{equation}}
\newcommand{\eeq}{\end{equation}}
\newcommand{\ben}{\begin{enumerate}}
\newcommand{\een}{\end{enumerate}}
\newcommand{\bit}{\begin{itemize}}
\newcommand{\eit}{\end{itemize}}
\newcommand{\barr}{\begin{array}}
\newcommand{\earr}{\end{array}}
\newcommand{\bc}{\begin{center}}
\newcommand{\ec}{\end{center}}
\newcommand{\DroII}{\overline{\overline{\rm D}}}
\newcommand{\DroI}{{\overline{\rm D}}}
\newcommand{\eps}{\epsilon}
\newcommand{\veps}{\varepsilon}
\newcommand{\vepsdi}{{\cal E}^\mathrm{d}_\mathrm{i}}
\newcommand{\vepsde}{{\cal E}^\mathrm{d}_\mathrm{e}}
\newcommand{\lraS}{\longmapsto}
\newcommand{\lra}{\longrightarrow}
\newcommand{\LRA}{\Longrightarrow}
\newcommand{\Equival}{\Longleftrightarrow}
\newcommand{\DRA}{\Downarrow}
\newcommand{\LLRA}{\Longleftrightarrow}
\newcommand{\diver}{\mbox{\,div}}
\newcommand{\grad}{\mbox{\,grad}}
\newcommand{\cd}{\!\cdot\!}
\newcommand{\Msun}{{\,{\cal M}_{\odot}}}
\newcommand{\Mstar}{{\,{\cal M}_{\star}}}
\newcommand{\Mdot}{{\,\dot{\cal M}}}
\newcommand{\ds}{ds}
\newcommand{\dt}{dt}
\newcommand{\dx}{dx}
\newcommand{\dr}{dr}
\newcommand{\dth}{d\theta}
\newcommand{\dphi}{d\phi}

\newcommand{\pt}{\frac{\partial}{\partial t}}
\newcommand{\pk}{\frac{\partial}{\partial x^k}}
\newcommand{\pj}{\frac{\partial}{\partial x^j}}
\newcommand{\pmu}{\frac{\partial}{\partial x^\mu}}
\newcommand{\pr}{\frac{\partial}{\partial r}}
\newcommand{\pth}{\frac{\partial}{\partial \theta}}
\newcommand{\pR}{\frac{\partial}{\partial R}}
\newcommand{\pZ}{\frac{\partial}{\partial Z}}
\newcommand{\pphi}{\frac{\partial}{\partial \phi}}

\newcommand{\vadve}{v^k-\frac{1}{\alpha}\beta^k}
\newcommand{\vadv}{v_{Adv}^k}
\newcommand{\dv}{\sqrt{-g}}
\newcommand{\fdv}{\frac{1}{\dv}}
\newcommand{\dvr}{{\tilde{\rho}}^2\sin\theta}
\newcommand{\dvt}{{\tilde{\rho}}\sin\theta}
\newcommand{\dvrss}{r^2\sin\theta}
\newcommand{\dvtss}{r\sin\theta}
\newcommand{\dd}{\sqrt{\gamma}}
\newcommand{\ddw}{\tilde{\rho}^2\sin\theta}
\newcommand{\mbh}{M_{BH}}
\newcommand{\dualf}{\!\!\!\!\left.\right.^\ast\!\! F}
\newcommand{\cdt}{\frac{1}{\dv}\pt}
\newcommand{\cdr}{\frac{1}{\dv}\pr}
\newcommand{\cdth}{\frac{1}{\dv}\pth}
\newcommand{\cdk}{\frac{1}{\dv}\pk}
\newcommand{\cdj}{\frac{1}{\dv}\pj}
\newcommand{\rad}{\;r\! a\! d\;}
\newcommand{\half}{\frac{1}{2}}

 \title{Glitching pulsars: unraveling the interactions of general relativity with quantum fields in the strong field regimes}

\author{A. A.~Hujeirat}
\email{AHujeirat@uni-hd.de}
\affiliation{IWR, Universit\"at Heidelberg, 69120 Heidelberg, Germany}
\author{R.~Samtaney}
\email{Ravi.Samtaney@kaust.edu.sa}
\affiliation{Applied Mathematics and Computational Science,
King Abdullah University of Science and Technology, Thuwal 23955-6900, Saudi Arabia}

\begin{abstract}{
 We present a  modification of  our previous model for the mechanisms underlying the glitch phenomena in pulsars and young neutron stars.
 Accordingly, pulsars are born with embryonic cores comprising of purely incompressible superconducting gluon-quark
  superfluid (henceforth SuSu-cores). As the ambient medium cools and spins down due to emission of magnetic dipole radiation,
  the mass and size of SuSu-cores are set to grow discretely with time, in accordance with the Onsager-Feynmann analysis of superfluidity.
  Presently, we propose that  the spacetime embedding  glitching pulsars is  dynamical and of bimetric nature: inside SuSu-cores the spacetime
  must be flat, whereas  the surrounding region, where the matter is compressible and dissipative, the spacetime  is Schwarzschild.
  It is further proposed that the topological change of spacetime is derived by the strong nuclear force, whose operating length scales
  is found to  increase  with time to reach $\mathcal{O}(1)~cm$ at the end of the luminous lifetimes of pulsars.
  The model presented here model is in line with the recent radio and GW observations of pulsars and NSs.
    }
\end{abstract}

\maketitle


 \section{Internal structure of UCSs}
  Pulsars, neutron stars (NSs) and magnetars build the family  of  ultra-compact stars (UCSs) with compactness parameter,  $ \alpha_S,$ generally  larger than half. Recalling that the average density of matter in the interiors of UCSs  is larger than the nuclear one, $\rho_0,$  then these objects are well-suited
  for probing the  interaction of general  relativity with quantum fields in the strong field regime.
\begin{figure}[htb]
\centering {\hspace*{-0.35cm}
\includegraphics*[angle=-0, width=7.0cm]{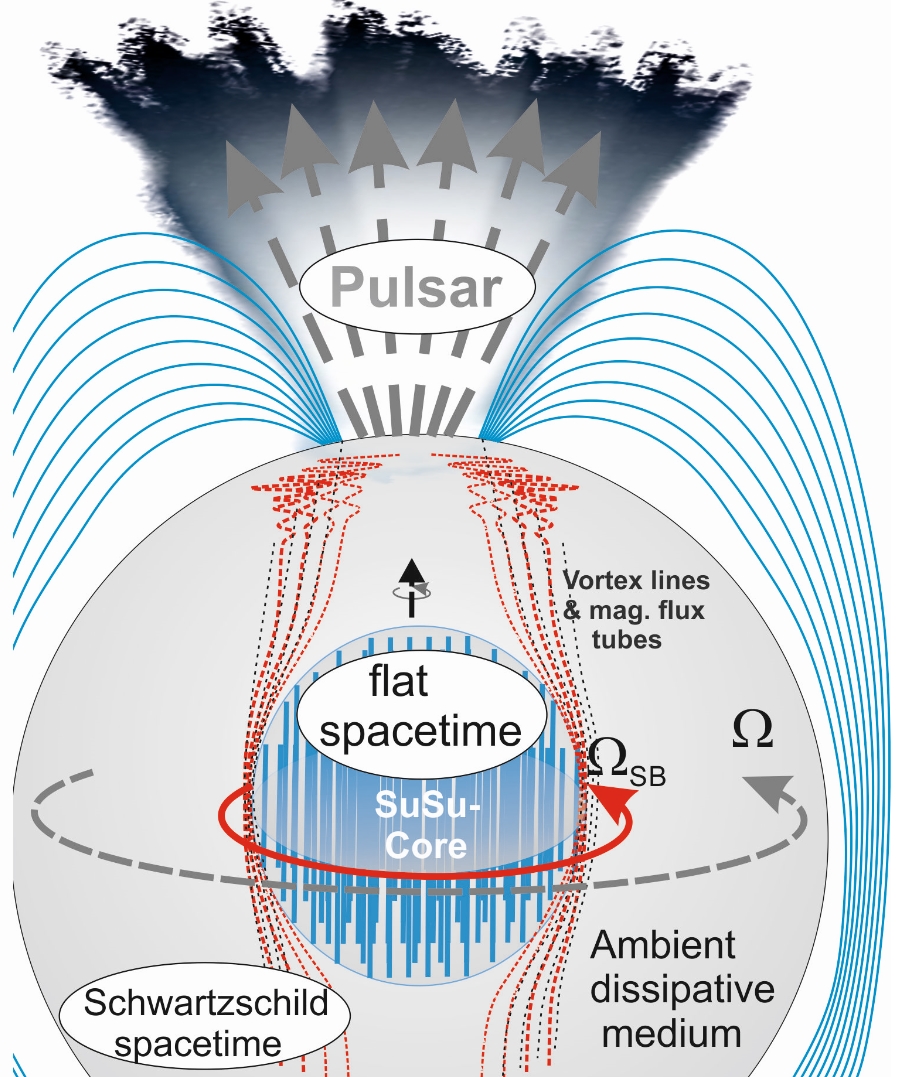}
}
\caption{\small Schematic depiction of the internal structure of glitching pulsars. Pulsars are born with  embryonic cores are comprised of an incompressible gluon-quark superfluid and embedded in flat spacetime. The overlying shells are made of dissipative and
compressible quantum fluid, but embedded in Schwarzschild spacetime.
}\label{NSInternal}
\end{figure}
  In fact, pulsars and NSs have been successfully used for probing general relativity, such as the emission of gravitational waves (GWs) in binary
  pulsars and  direct detection GWs through merger of neutrons stars and black holes (see  \cite{NSMergerLIGO} and \cite{GWReview19} for
  detailed reviews). However, the internal structure of UCSs continues to be a mystery and
  remains a challenge for astrophysicists (see the review in \cite{ReviewPulsars16} and the references therein).  On the other hand, the observed glitch-phenomena associated with pulsars are events that may be considered to be connected to the dynamics of matter inside their interiors (see \cite{ Cook1994,Camenzind2007, Espinoza2011, ReviewPulsars16} and the references therein).
  However, based on observations and theoretical arguments,  one may try to constrain the nature of matter inside normal UCSs
   as follows:
\begin{enumerate}
\item  The observed spin-down of UCSs results from the loss of magnetic and rotational energies that are estimated to be of order  $10^{38}$ erg/s. This  would imply complete exhaustion of the stored  removable energies\footnote{ i.e. energies other than rest energy} inside pulsars within about ten million years, provided that the heat conductivity operates on length scales that are larger than the nuclear ones. Consequently, very old and isolated  UCSs, including those formed through the collapse of the first generation of stars,  must be by now dark and invisible and therefore excellent black  hole candidates.
This argument would explain  why neither UCSs nor black holes  in the mass range   $[2\MSun \leq \calM \leq 5 \MSun]$ have ever been observed.
 \item The rest of thermal, kinetic and magnetic energies that are left from the collapse of the
 progenitors of UCSs are transported outwards into the outer shells and subsequently liberated away.
 In the absence of nuclear energy generation, the  Tolman-Oppenheimer-Volkoff  equation (TOV-equation)  may still accept
a positive gradient of the thermal energy, turning the central core to be the coldest region inside the object and thereby facilitating a phase transition  of the compressible dissipative nuclear matter into incompressible superfluid.  Such a transition may still occur even when the matter's temperature is still beyond several  million degrees.
\item Theoretical studies show that in the regime of nuclear density and beyond, almost all EOSs used for modeling state of matter in the  interiors of UCSs  tend to converge to the limiting case $ \varepsilon = P$ (see \cite{Camenzind2007} and the references therein). However, this  state corresponds to pure incompressible fluids \cite{HujeiratMassiveNSs18}.  Due to causality and stability reasons, except rest energy, the incompressible fluid should be free of all other removable energies, such as thermal, kinetic and magnetic energies. Hence, the core  is expected to end up  as  a condensate with zero-entropy, whose constituents communicate with each other at the speed of light and therefore the momentum exchange between them must saturate around a universal maximum. In this case, the coupling constant in the context of asymptotic freedom in QCD would converge to its universal minimum value, where quarks should be moving freely throughout the whole domain \cite{Bethke2007,QCD18}. Such a zero-entropy superfluid\footnote{vanishing energy loss due to acceleration/deceleration should be excluded} cannot accept spatial variations or stratification due to gravity, and therefore  the
embedding spacetime must be flat. This state may be maintainable, if the involved quarks are attached to a spatially fixed lattice, where
they may behave collectively as a single macroscopic quantum entity.
 Indeed, although both entropy and energy states of colliding protons at the RHIC and LHC differ significantly from those at the
centers of UCSs,  these experiments  have shown that the properties of the resulting fluids mimic those of perfect fluids
(see \cite{LHCb2015,QGP2019} for further details).

\item {Based on astronomical observations, there appears to be a gab in the mass spectrum of ultra-compact objects: neither black holes
             nor NSs have ever been observed in the mass range $[ 2\,\mathcal{M}_\odot < \mathcal{M} < 5\,\mathcal{M}_\odot].$ Moreover,
             intensive astronomical observations of the  newly detected NS-merger GW170817 [1] failed to unambiguously classify
             whether the resulting object is a stellar black hole or a NS. These both facts are in line with our scenario that pulsars and
             young neutron stars evolve on the cosmological time scale into massive dark energy objects that become indistinguishable from stellar
            black holes.   A schematic description of the scenario is shown in Figs. (\ref{NSInternal},\,\ref{FeynmannDiagram}) and (\ref{CosmoTov}).  } 
            
\end{enumerate}

\begin{figure}[htb]
\centering {\hspace*{-0.35cm}
\includegraphics*[angle=-0, width=7.15cm]{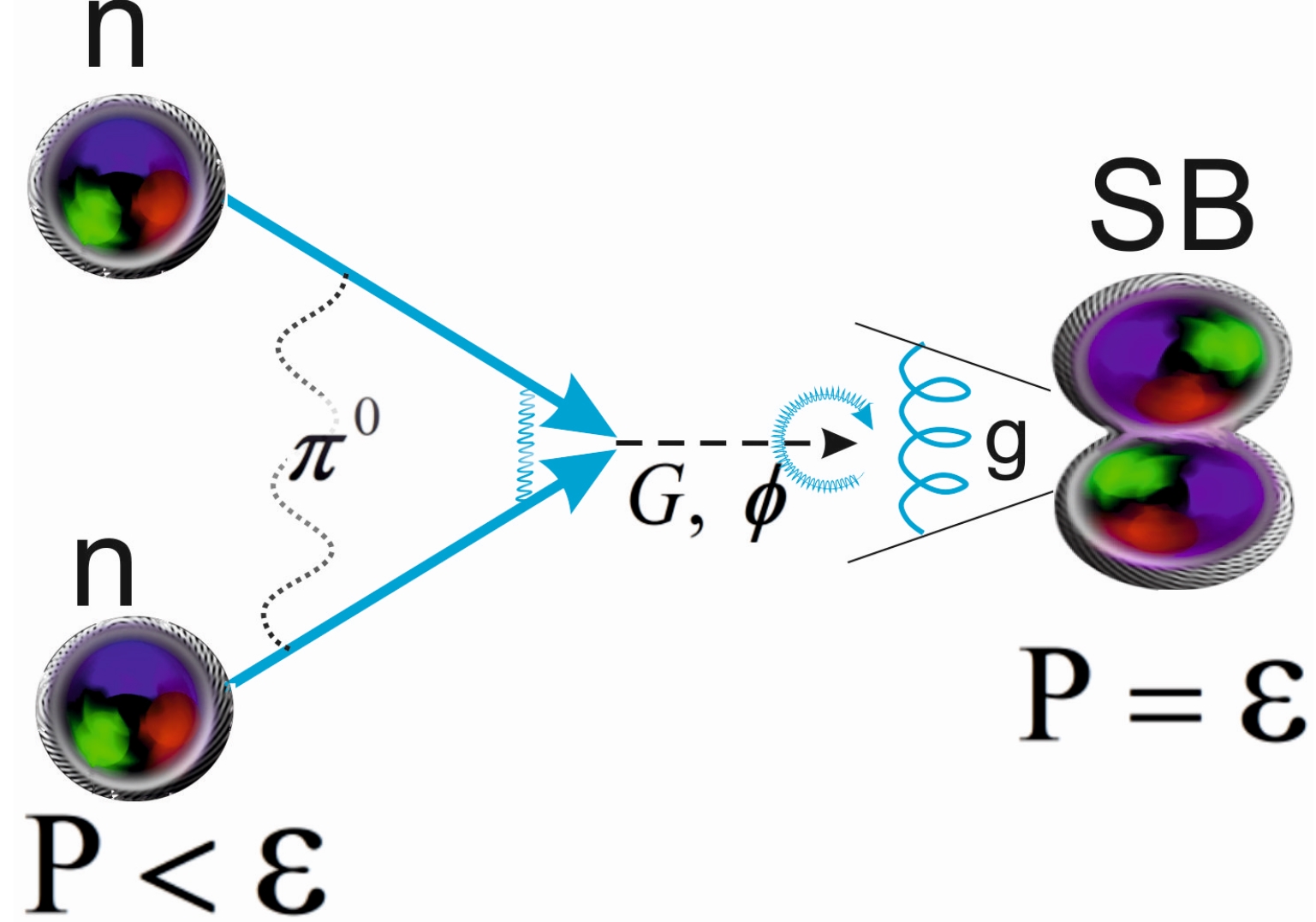}
}
\caption{\small  Neutron merger at the centers of pulsars. The compressible and dissipative neutron fluid is governed by  equations of state (EOSs) in which the pressure
$P < \varepsilon.$  When the density approaches  the critical density, $\rho_c,$ then the EOS changes into $P = \varepsilon,$ which
corresponds to incompressible quantum fluid. At $\rho=\rho_c,$ it is energetically favourable for neutrons to begin
deconfining the quarks and merge together to form an ocean of gluon-quark superfluid, though the resulting global
 gluon cloud must be much more energetic.
}  \label{FeynmannDiagram}
\end{figure}

\subsection{The model and the governing equations }
 For modeling the internal structure of pulsars, the general relativistic field equations should be solved;
  \beq
               G_{\mu\nu} = \kappa T_{\mu\nu}, \textrm{~~~~ for }  {\mu,\nu}:0\rightarrow 4,
  \eeq
  where $G_{\mu\nu},\, T_{\mu\nu}$ are the Einstein and the Stress-Energy tensors, respectively, and $\kappa$ is a coefficient.
 The rotational, magnetic, thermal and other energies in UCSs are generally about three orders of magnitude smaller than the rest energy and therefore they may be safely neglected.  In this case,  Schwarzschild spacetime is most suited for modelling their internal structures.  Hence, the  corresponding metric reads:
      \beq
   ds^2 = e^{2\nu(r)}dt^2 - {e^{2\lambda(r)}} dr^2 - r^2 d\Omega^2,
   \eeq
   where   $d\Omega = d\theta^2 + sin^2 \theta d\varphi^2$    is a surface element  on a sphere of radius ``r".
 The field equations can then be reduced to just one single equation, i.e.  the so-called  Tolman-Oppenheimer-Volkoff (TOV) equation:
   \beq
               \DD{dP}{dr} = -\DD{G}{c^4 r^2}\DD{[\varepsilon + P][m(r) + 4 \pi r^3 P]} {1 - r_s/r},
  \eeq
  where $r,\,r_s,\,G,\,P,\,m(r)$ correspond to the radius, Schwarzschild radius, gravitational constant, pressure and the dynamical mass, respectively
 (see  Sec. (4) in \cite{HujeiratMassiveNSs18} for further details).

Recalling that GR is incapable of modeling gravitationally bound incompressible matter, then the spacetime embedding pulsars
may be decomposed into two separate domains: a flat spacetime that embeds the SuSu-core and a surrounding Schwartzschild spacetime that embeds the ambient media as mentioned above. The decomposition of the domain is motivated by relativistic causality which prohibits fluid stratification or spatial variation of  the density of purely incompressible  fluids; hence the spacetime embedding the core must have zero-curvature, i.e. purely flat.

Based thereupon the line element in the core  is described by the metric in spherical polar  coordinates  $(t,\,r,\, \theta,\,\varphi)$ :
    \beq
   ds^2 = c^2 dt^2 - dr^2 - r^2 d\Omega^2.
   \eeq
 The physical properties of the SuSu-core are set to affect the structure of the ambient medium by  allowing $m(r)$  and $r_s$
in the TOV-equation   to depend on the total enclosed mass, i.e. on   $m(r) = M_{SuSu} + 4\pi \int^\infty_{r=r_{SuSu}} \rho r^2 dr.$ \\

On the other hand, as the matter in the rotating core is in incompressible superfluid state, it must contain a discrete array of vortices. Their total number, N,  the mass and size growth of the core are determined through the Onsager-Feynmann equation for modeling
quantized circulations:
 \beq
               \oint v\cdot d\ell = \DD{2\pi \hbar}{m^*} N,
 \eeq
where $v,\,\ell,\,\hbar,\,m^*,\,N$ correspond to rotational velocity, line-element along the circular path, the reduced Planck constant, the
mass of the superfluid particles and the enclosed number of vortices, respectively \cite{Yarmchuk1979, SFluidTurbo2007}.

Inside the core, the quantum fluid  is assumed to have zero-entropy and has reached the critical
supranuclear density, $\rho = 3\, \rho_0, $ beyond which merger of individual gluon clouds into a global one becomes possible.
The effective energy of the latter cloud correlates linearly with the number of merged neutrons, i.e.
      \beq
    \sum_1^N  E^0_n   \xrightarrow[ ]{mergering ~ process} \sum_1^{N} E^0_n + N\times \Delta E_{bag} ,
   \eeq
where
$ \Delta E_{bag} $ is the bag energy enhancement needed to confine the quarks inside the super-baryon and
$N$ denotes the total number of merged neutrons to forming the super-baryon, $E_n$ is the rest energy of a single neutron.
In this case  the energy and pressure of the incompressible gluon-quark superfluid inside the core are as follows:
   $\varepsilon_{tot} =   \varepsilon_0 +  \varepsilon_{\phi}, ~~   P_{tot} =   P_0 +  P_{\phi}, $
   where $\varepsilon_0$ is the energy density of the baryonic matter and
   $\varepsilon_{\phi} =\DD{1}{2} \dot{\phi}^2  + V(\phi) + \DD{1}{2} (\nabla \phi)^2$ and
   $P_{\phi} =\DD{1}{2} \dot{\phi}^2  - V(\phi) + \DD{1}{6} (\nabla \phi)^2$ are the energy density and pressure of the  scalar field,
   which are assumed to be identical to $\Delta E_{bag}.$
 As the matter inside the cores of pulsars is set to be incompressible and stationary, it is reasonable to assume   $\phi= constant.$ In this case,
$\dot{\phi}$ and $ \nabla \phi$   must vanish and  $V(\phi) $ can be considered  as the energy density required for deconfining the quarks inside the super-baryon, i.e., the SuSu-core.  Here we set $\varepsilon_{tot} =  2 \varepsilon_0,$ whereas the pressure $P_{tot} = P_0 - V(\phi) = \varepsilon_0 - V(\phi) = 0.$ Thus, the total local pressure of  incompressible gluon-quark superfluid inside SuSu-cores vanishes completely.
Exterior to the core, the matter is said to be baryonic, dissipative, compressible and embedded in a Schwarzschild spacetime.


\subsubsection{ Initial and boundary conditions}
The present scenario is based on a previous model for the origin of glitches in pulsars.
Accordingly, pulsars and young neutron stars are expected to undergo billions of glitch events during their luminous lifetimes before ending as ultra-compact and invisible  dark energy objects.
In the present study, we select  several  epochs in the lifetimes of pulsars  and the corresponding internal structures are
calculated subject to the following conditions:
    \begin{description}
    \item[$\blacklozenge$] The pulsar has the  initial mass  of $1.33 \mathcal{M}_\odot,$ made of a purely baryonic compressible matter.
    \item[$\blacklozenge$]  The pulsar is set to have initially the compactness parameter $\alpha_s=1/2.$
    \item[$\blacklozenge$]   The central baryonic density is set to be  equal to the critical $\rho_c= 3\, \rho_0,$  at which a transition into
              quark deconfinement occurs, and where the Gibbs function  vanishes.
    \item[$\blacklozenge$]   The dissipative and compressible baryonic matter is set to obey the polytropic EOS:  $ P = \mathcal{K}$  $ \rho ^\gamma.$
\end{description}

   The selected models  correspond to  pulsar phases in which the enclosed SuSu-cores have reached the following  radii: $R_{SuSu}= \,0.333,\, 0.525,\, 0.78525 \textrm{~and~} 0.8575  $ in units of  $[\tilde{R} ~(= R_S/\alpha_s)],$ where $R_S$ i­s the corresponding Schwarzschild radius.
   The total density inside SuSu-cores  amounts to:  $ \rho_{tot} = \rho_{b}+\rho_{\phi}=  2\, \rho_c \approx 6\,\rho_0.$
      \ben
       \item  Model-0: The pulsar is made of purely baryonic matter.    Here,   the  TOV-equation is solved for the pressure
              starting from a given $P(r=0)=P_0 $ up to a radius, where the pressure vanishes.
       \item Combined-models:  The pulsar models are made both of SuSu-cores surrounded by dissipative and compressible quantum
                fluid.The radii of the cores are:  $R_{SuSu}= 0.333,\, 0.525,\, 0.78525, 0.8575.$
                The total  mass of the core is calculated through the integration:
              \beq
            m_{tot}(r=R_{core}) = \int_{0}^{R_{core}} (\varepsilon_b + \varepsilon_\phi) dr.
            \eeq
            The solutions here are based on adapting the parameters $\mathcal{K}$ and $\gamma$ of the EOS in a manner such that
            the initial baryonic mass of $1.33~\mathcal{M}_\odot$ the pulsar  remains preserved.
       \item The ultimate final phase: Here the radius and mass of the SuSu-core are roughly equal to the critical values:
       $R_{SuSu}= 0.859 \approx R_{Schw.}$ and $M_{SuSu} \approx M_{Schw.}.$  Here the initial pulsar must have metamorphosed
       entirely into  an invisible SuSu-object.
      \een
\subsubsection{Results }
\begin{figure}[htb]
\centering {\hspace*{-0.35cm}
\includegraphics*[angle=-0, width=7.5cm]{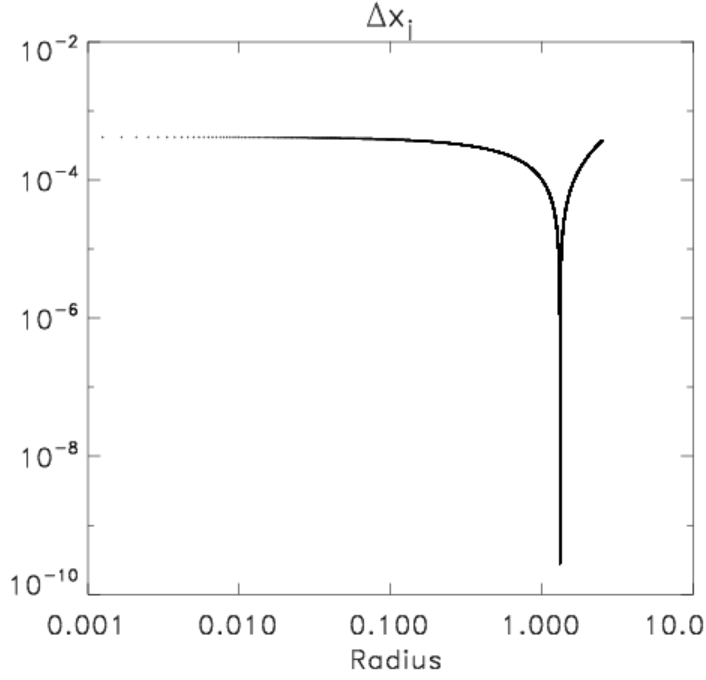}
}
\caption{\small  The distribution of the finite volume cells as function of radius. Here the explicit adaptive mesh refinement (EAMR)
has been employed to increase the accuracy at both the interface between the core and the ambient medium as well as to outer radius of the object.
}\label{GridDistribution}
\end{figure}

For enhancing the spatial accuracy of the calculations, an explicit adaptive mesh refinement (EAMR) has been developed, in which the aspect ratio,
$dr_{max}/dr_{min},$ may reach  100 million. Unlike dynamic adaptive mesh refinement (AMR),   EAMR is based on {\it a posteriori}
refining the grid distribution in certain locations of the domain, where the gradients of the physical variables are large. This may be achieved
by manually manipulating the grid distribution and restarting the calculations anew.
In the present case, for example, the location where the  pressure vanishes is vitally important and therefore the resolution
should be maximally refined (Fig.\ref{GridDistribution}).  Unlike AMR,  solution methods based on EAMR generally converge much faster than
their AMR-counterparts, hence contributing to the global efficiency and robustness of the numerical solution procedure (see \cite{FischerHuj2019} and the references therein).\\

\begin{figure}[htb]
\centering {\hspace*{-0.35cm}
\includegraphics*[angle=-0, width=7.5cm]{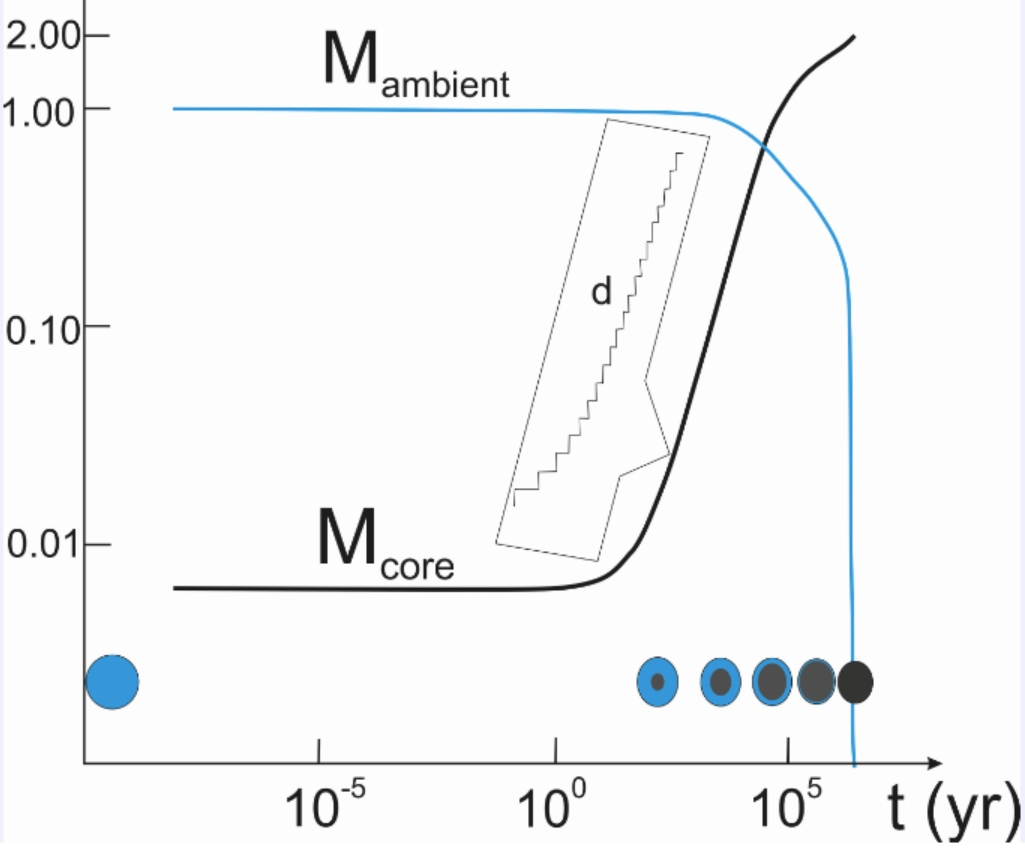}
}
\caption{\small Based on the solution of the TOV equation in combination with Onsager-Feynmann equation, the
size and mass of SuSu-cores grow with time following a well-defined  mathematical sequence    $\{\alpha^n_c\}.$
The discrete increase of $\mathcal{M}_{core}$  is magnified and shown  ``d".
To verify the bimetric model of pulsars, six epochs with different core-sizes have been selected: a newly born pulsar, four intermediate
phases and the final massive state, where the whole object turns into an invisible dark energy object.
}\label{CosmoTov}
\end{figure}

\begin{figure}[htb]
\centering {\hspace*{-0.35cm}
\includegraphics*[angle=-0, width=7.5cm]{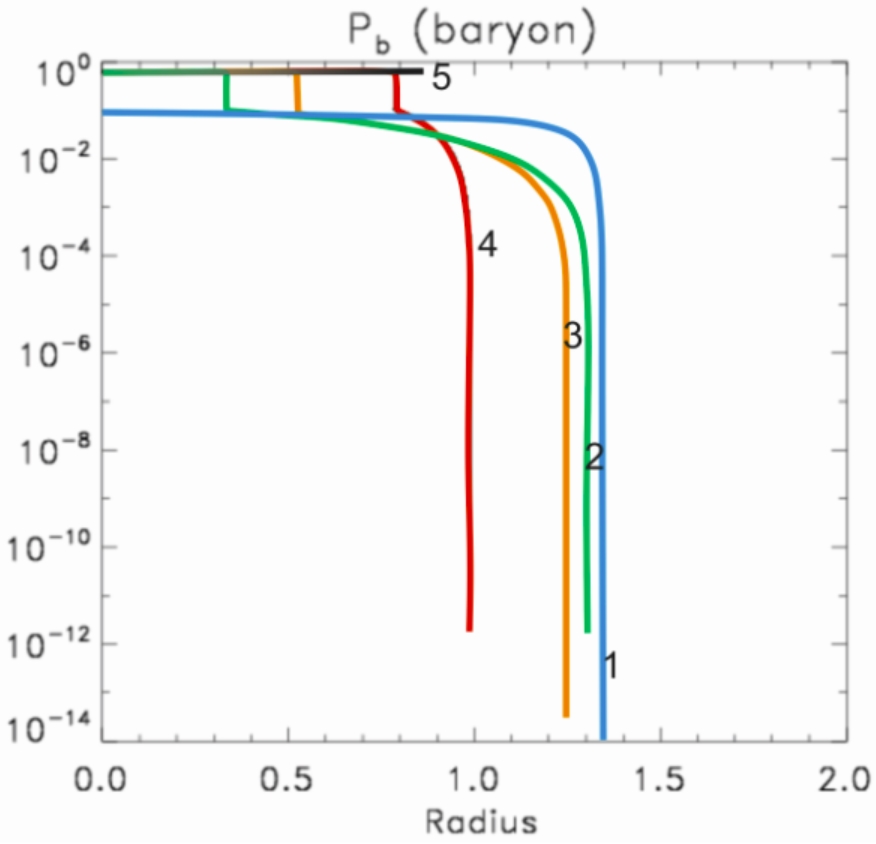}
}
\caption{\small The distribution of the baryonic pressure inside an evolving pulsar after five selected glitch events at different epochs.
Inside the core, in addition to the rest energy of the baryonic matter,  there is  an energy enhancement due to the scalar field, which is equivalent
to the additional energy required for re-confining  the sea of quarks. The transition between both regions
is not smooth as the matter inside the core evolves quantum mechanically,  whereas the ambient medium obeys the normal laws of continuum.
}\label{PressureMany}
\end{figure}

\begin{figure}[htb]
\centering {\hspace*{-0.35cm}
\includegraphics*[angle=-0, width=7.5cm]{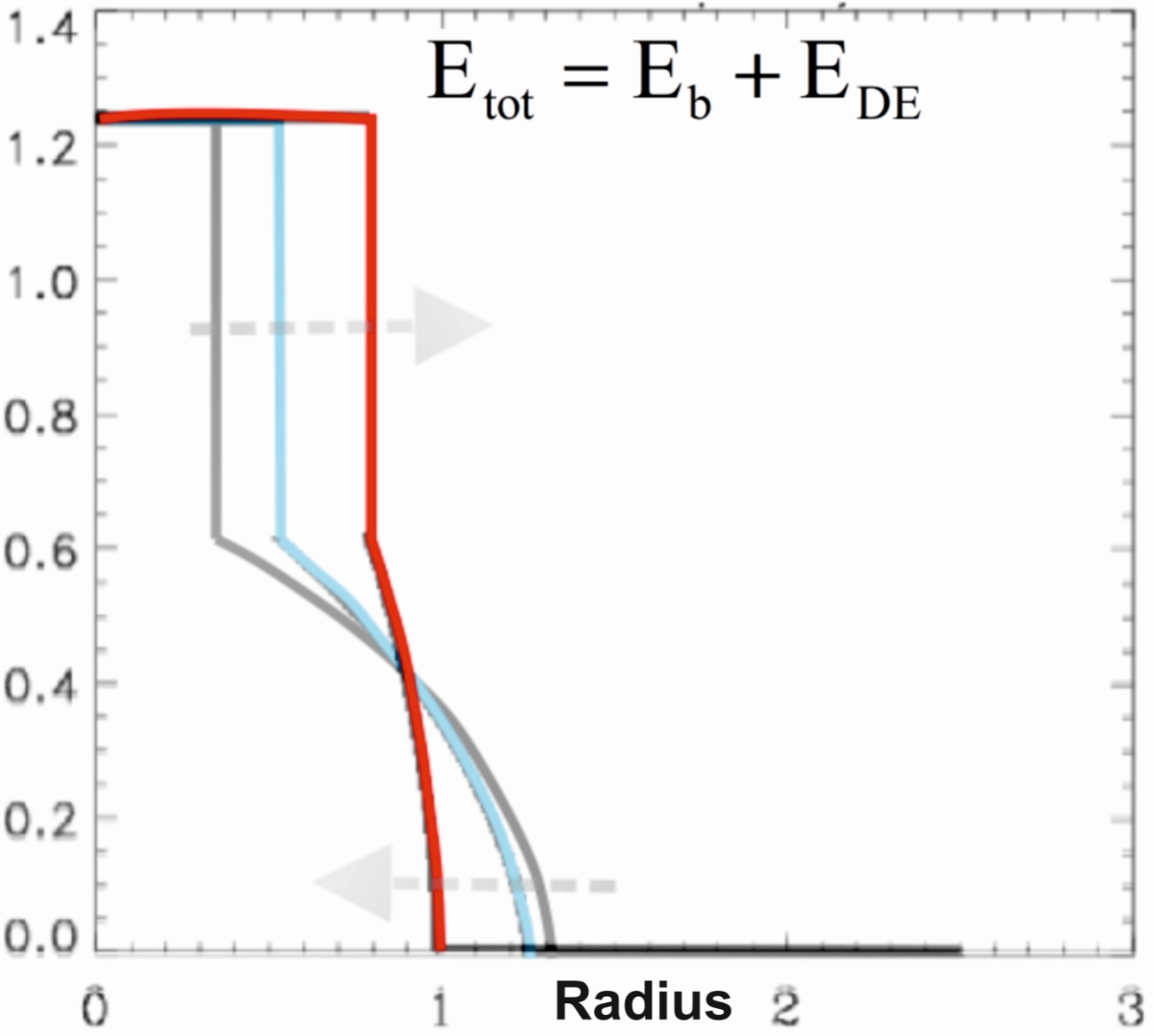}
}
\caption{\small  Similar to the previous figure: the distribution of the total energy density inside both the core and the ambient dissipative medium
are shown.
}\label{EnergyMany}
\end{figure}

\begin{figure}[htb]
\centering {\hspace*{-0.35cm}
\includegraphics*[angle=-0, width=7.5cm]{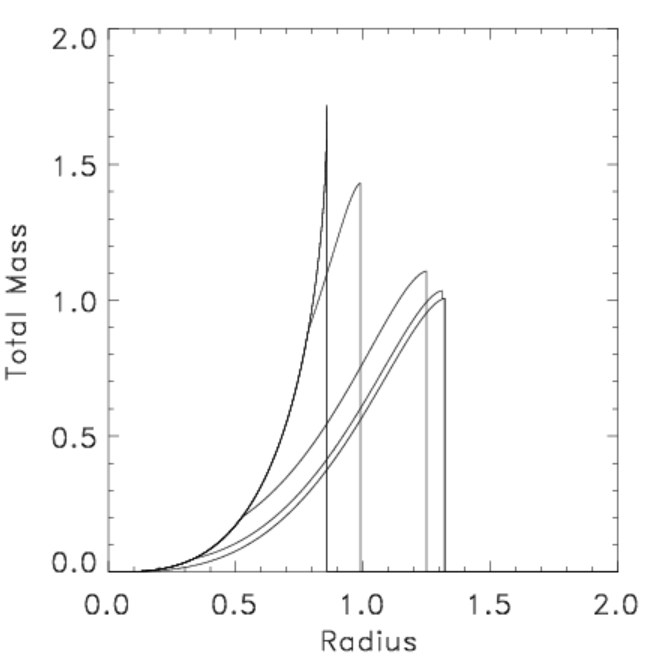}
}
\caption{\small The total enclosed mass of the object versus radius after selected glitch events. As the mass of the SuSu-core
increases, its compactness of the whole object  increases as well to finally reach the critical Schwarzschild mass.
}\label{MassIncrease}
\end{figure}

\begin{figure}[htb]
\centering {\hspace*{-0.35cm}
\includegraphics*[angle=-0, width=7.5cm]{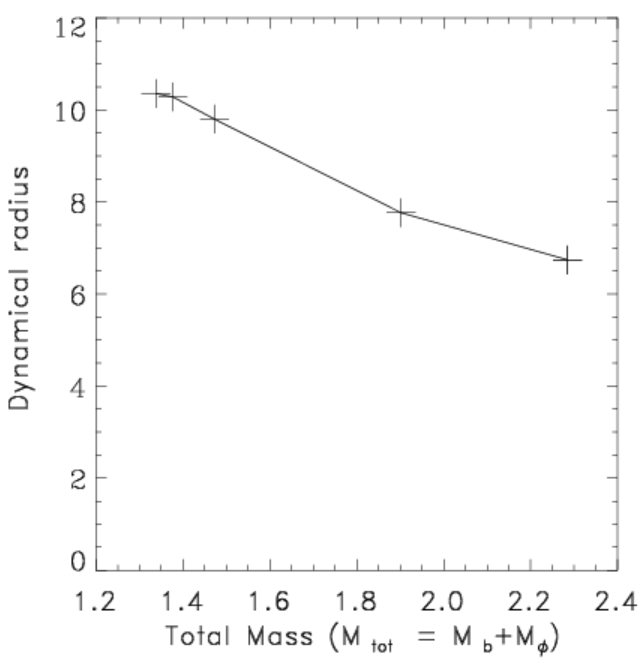}
}
\caption{\small The mass-radius relation of an evolving pulsar after selected glitch events. As the core becomes more massive,
the Schwarzschild radius grows and converges asymptotically to the effective radius of the entire contracting object.
}\label{MassRadiusRelation}
\end{figure}

{Based on the previous study \cite{HujeiratGlitch18}, five episodes in the cosmological evolution of pulsars have been selected
 (see Fig. \ref{CosmoTov}). For each episode, the TOV equation, modified to include dark energy input at the background
 of the here-presented bimetric scenario of spacetime, has been  solved. \\
 In Figs. (\ref{PressureMany}, \ref{EnergyMany}) the profiled of the corresponding five runs are displayed.  }
In Figs.~\ref{PressureMany} and~\ref{EnergyMany}) show the profile of pressure and total energy density, respectively, for five different runs.
 These are based on the aforementioned bimetric scenario
of spacetime. Profile (1) corresponds to the radial distribution of the pressure obtained by solving the TOV equation.  The phase here corresponds
to the moment of birth of the pulsar, when it is entirely  embedded in a Schwartzschild spacetime.
Profiles (2, 3, 4 and 5) in Fig.(\ref{PressureMany}) show the radial distributions of the pressure at different epochs, specifically when the SuSu-core has grown in mass and size
to reach the sequence of radii: $R_{SB}= 0.333,~0.525,~0.78525~\textrm{and}~0.7875.$
Note that as the SuSu-core becomes more massive,  the curvature of spacetime embedding the ambient medium is enhanced, which, in turn,
 compresses the ambient medium even more, thereby reducing the effective radius of the pulsar. This would explain the mass-radius relations
 displayed in Figs. \ref{MassIncrease} and ~\ref{MassRadiusRelation}. Here the corresponding Schwarzschild radius increases and propagates outwards
 to finally meet the decreasing  radius of the contacting pulsar. The overlapping of the both radii is expected to occur at the end of the pulsar's luminous lifetime (see Fig. \ref{CosmoTov} and Profile 5 in Figs.(\ref{PressureMany}) ).

\begin{figure}[htb]
\centering {\hspace*{-0.5cm}
\includegraphics*[angle=-0, width=7.85cm]{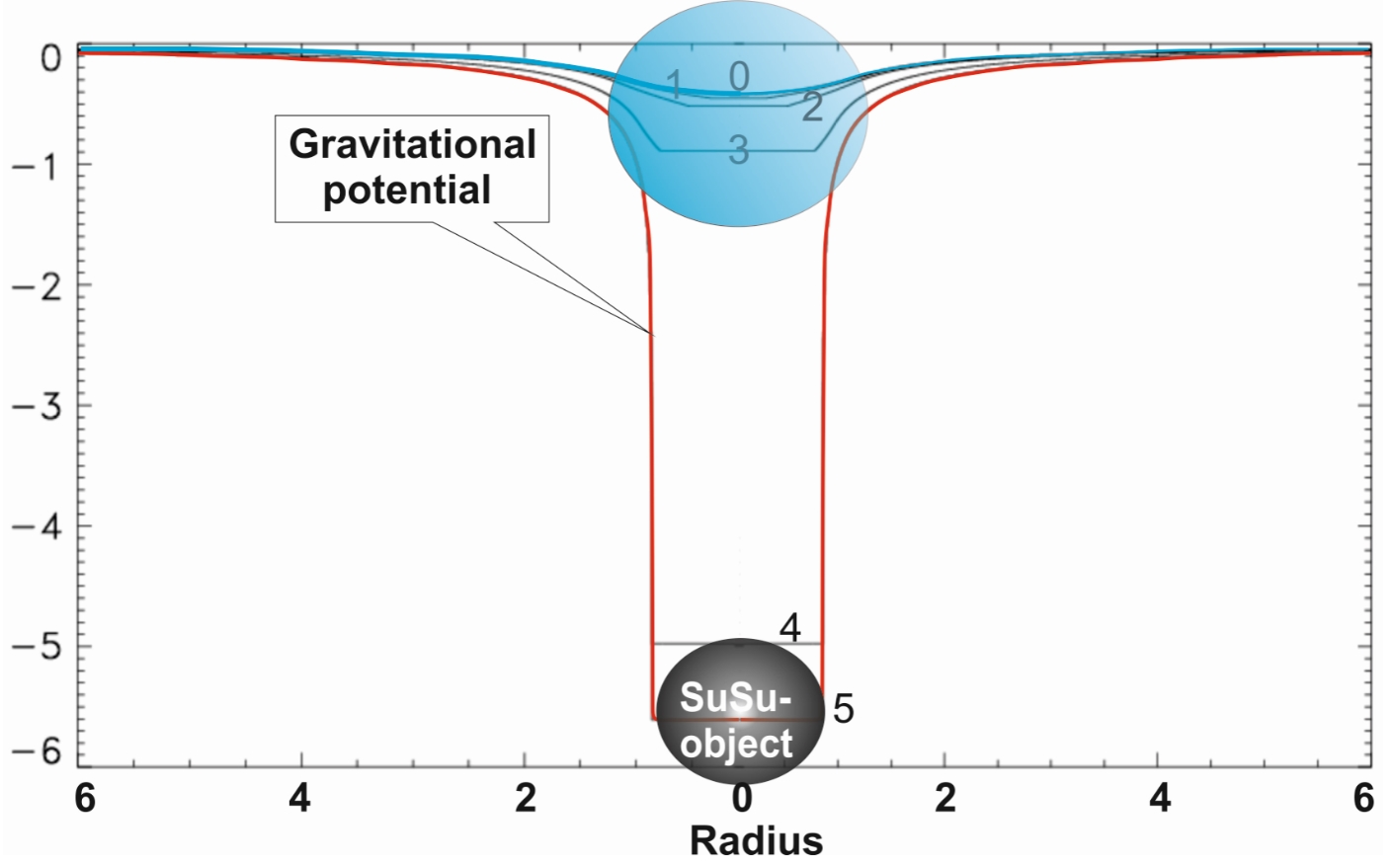}
}
\caption{\small  The radial distribution of the gravitational potential shown for different epochs in the lifetime of the pulsar, which is equivalent to the radial projection of spacetime
 inside the SuSu-core governed by Minkowski spacetime,  as well as in the surrounding region governed by the Schwarzschild metric. Obviously, as
the object become more massive, it sinks deeper in spacetime, becomes gravitational redshifted to finally ends as an invisible
dark energy object.
}\label{GravPotential}
\end{figure}

\begin{figure}[htb]
\centering {\hspace*{-0.5cm}
\includegraphics*[angle=-0, width=8.5cm]{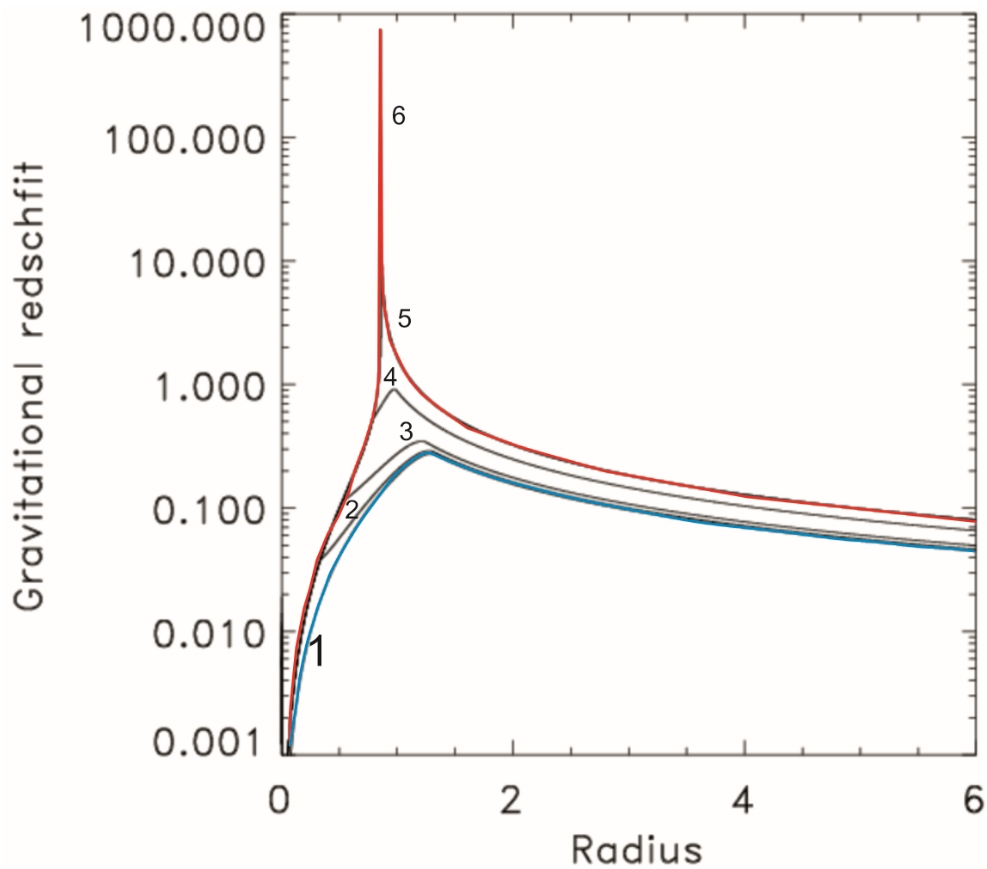}
}
\caption{\small The radial distribution of the gravitational redshift (Z) of the pulsar at different evolutionary epochs.  Profile ``1" corresponds to the low
``Z" immediately after the pulsar was born, whereas Profile ``6"  depicts the divergent limit of  ``Z"  at the surface of the object when it becomes
completely invisible.
}\label{Redschift}
\end{figure}

As purely incompressible superfluids in flat spacetime have zero-spatial variations (see Fig.(\ref{GravPotential}),
 then the gravitational potential inside SuSu-cores should attain a sequence of constant values. Their magnitudes
  correlate with  the mass and size of the SuSu-core. As a consequence, as the SuSu-core becomes more massive,
  the gravitational redshift of the pulsar increases as well to finally reach very large values at the end of the pulsar's luminous
  lifetime  (see Profile ``6" in Fig.\ref{Redschift}). Here the Schwarzschild radius becomes almost equal to the effective radius of the
object, enforcing the object to sink deeply in spacetime and becomes invisible.\\
Note that the profiles shown in Figs.(\ref{GravPotential} and  \ref{Redschift}) are not just schematic representations, but rather have been obtained
 using direct numerical computations.



\section{Summary and discussion}
The pulsar model presented here is analogous to a terrestrial experiment of a  rotating superfluid Helium in a container.
While  the  SuSu-matter inside the  core corresponds to  a rotating superfluid Helium, the dissipative ambient medium
would be represented by the solid container.
  However, the pulsar model  here is much more sophisticated, as quantum and general relativistic effects are taken into account.
   In the following we discuss the main features of the model:
   \begin{figure*}
\centering {\hspace*{-0.35cm}  \vspace*{0.35cm}
\includegraphics*[angle=-0, width=16.5cm]{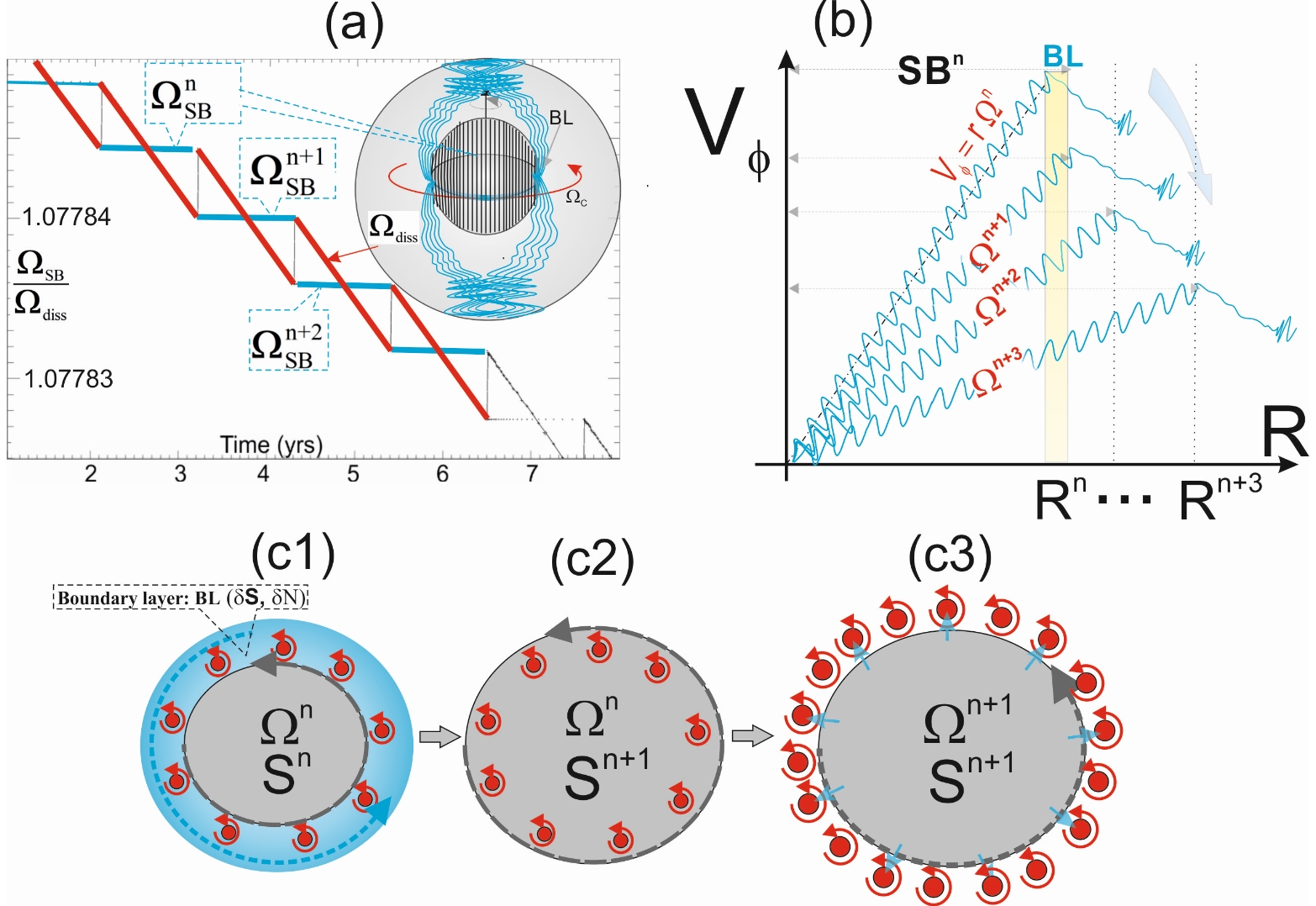}
}
\caption{\small   The three-stage glitch scenario. In (a) the time-development of both
rotational frequencies of the SuSu-core (blue-color) and of the ambient medium (red-color) are shown, using
GR-numerical calculations. In (b) the rigid-body rotation and increasing size of the SuSu-core for four successive glitch-events
are shown. The boundary layer here has the width $\delta R,$ area $\delta S$  and contains $\delta N$ vortices.
In the lower panel we show the three-stage scenario of the glitch phenomena in pulsars. In (c1) the SuSu-core and
boundary at the verge of a glitch event are shown. Once the SNF has locked the neutrons inside the BL,
 to the core and enforced them to adopt the same thermo- and hydrodynamical properties of the matter inside the core,
 then the spacetime embedding  the BL undergoes a topological change into a flat spacetime and merges with that of the core (c2).
However, once the resulting difference $\Delta \Omega$ between $\Omega^n_{SB}$ of the core and the ambient medium $\Omega^n_{AM}$
has surpassed a critical value, then the core must undergo a transition into the next  lower energy state by expelling a certain number of vortices (c3).
In turn, the ambient medium absorbs the vortices and  re-distribute them viscously, thereby giving rise to the prompt spin-up observed in gliching pulsars.
}\label{3StageScenario}
\end{figure*}
\begin{itemize}
  \item For $\rho \geq \rho_{crit},$ neutrons at the center of pulsars are set to merge together to form an incompressible gluon-quark superfluid, governed by the     EOS $ P = \varepsilon.$  This argument is supported by the following observations:
           \begin{description}
             \item[$\lozenge$]  Theoretical studies of nuclear interactions in the regime $\rho \geq \rho_0$  show that almost all EOSs converge to
             the limiting case: $ P = \varepsilon=  a_0 n ^2,$ which corresponds to the pure incompressible state. However, once the matter is governed by
                $P~=~\varepsilon,$ the chemical potential $\mu$ would achieve a universal maximum, i.e. $\DD{\D \mu}{\D n} =0.$
               \item[$\lozenge$] based on thermodynamical considerations, it was shown in \cite{HujeiratMassiveNSs18} that  at $\rho = \rho_{crit} \approx 3 \times  \rho_0,$
               in combination with zero-entropy condition,  the Gibbs function attains a global zero-minimum, facilitating thereby a crossover phase transition into a globally confined gluon-quark superfluid, where $\DD{\D \mu}{\D n} =0$.
             \end{description}
  \item The formation of incompressible gluon-quark superfluids is associated with energy enhancement of the gluon field at the surface
            of the core, which is  needed for globally re-confining the enclosed quarks,  thereby effectively  enhancing the effective mass  of the core.
  \item The phase transition of the quantum fluid in the boundary layer (BL) from compressible-dissipative into incompressible superfluid states is associated with changes
           of the spacetime topology from a curved spacetime into a purely flat one. The latter reaction is due to causality, which prohibits
           incompressible superfluids to be embedded by a curved spacetime.
           Indeed, the topology change of spacetime is associated with the emission of gravitational waves. However, due to the tiny little volume
           of the BL, the detection of GW-emission during glitch-events would be much below the sensitivity of today GW-detectors.

  \item Both the phase transition of matter and the change of topology of spacetime are provoked by the SNF, which appear to develops into a macroscopic force   as the pulsars ages.
            In fact, as the core is made of SuSu-matter, the core is equivalent to a super-baryon, in which the enclosed quarks are shielded by
            a gluon cloud \cite{Witten1984}. As will be shown later, the SNF transmitted  by this cloud is found to correlate nicely with the size and mass of the core.
            The neutrons in the BL become increasingly locked to the core, thereby  adapting the physical conditions of the matter  inside the core.
            Following the scenario of \cite{HujeiratGlitch18}, the occurrence of glitches in pulsars follows a well defined sequence, whose elements are
             $\{\alpha^n_c\}$ where
             $\alpha^n_c =\Omega^{n}_{c}/\Omega^{n+1}_{c} = 1 + (\Delta\Omega/\Omega)^n \}, $ and $n: 0\rightarrow \infty.$
             Here $n=0$ corresponds to the first glitch event immediately after the pulsar's formation, whereas  $n=\infty$ corresponds to
             the final glitch event at the end of the pulsar's luminous lifetime. \\
             Thereupon,  let $R_n, \Omega_n$ be the radius and angular frequency of the
             SuSu-core on the verge of the glitch event number $"n,$ respectively. The
            rotational energies of the SuSu-core at time $t_n$ and $t_{n+1}$ read:
 \begin{equation*}
            \begin{array}{l}
             E^n_{rot,SB} = \DD{4 \pi \rho_{crit}}{15} R^5_n \Omega^2_n \\
             E^{n+1}_{rot,SB} = \DD{4\pi\rho_{crit}}{15} R^5_{n+1} \Omega^2_{n+1}.
             \end{array}
 \end{equation*}

            From conservation of rotational energy:  $E^n_{rot,SB}=E^{n+1}_{rot,SB}, $ we obtain the relation:
            \beq
            (\DD{R_{n+1}}{R_n})^5  = (\DD{\Omega_n}{\Omega_{n+1}})^2.
            \eeq
            Inserting  $R_{n+1}= R_{n} + \delta R^{BL}_n , $  $\Omega_{n}= \Omega_{n+1} + \delta \Omega_n  $ and using Taylor-expansion, then
            the width of the BL is:
              \beq
               \DD{\delta R^{BL}_n}{R_n} \approx \DD{2}{5} \DD{\delta \Omega_n}{\Omega_n}.
              \eeq
            $\delta R^{BL}_n, \delta \Omega_n$ here  denote the width of the boundary layer between the SuSu-core and the ambient dissipative medium and the absolute difference between the rotational frequencies of the core before and after the glitch event, respectively.\\
            Based thereon,  the width of the BL during different evolutionary epochs of the pulsar may be estimated as follows:
        \beq
        \hspace*{-2mm}
     \DD{\delta R^n_{BL}}{R^n_{SB}}\approx
      \left\{
       \begin{array}{ll}
               1.4 \times  10^{-10} & \tau_{age}=0 \\
               2.2 \times 10^{-8} &  \tau_{age} =~ 1000~ yrs/Crab \\
               1.6 \times 10^{-6} &  \tau_{age} =~ 10000~ yrs/Vela\\
               3.8 \times 10^{-6} & \tau_{age}~ =~ 10\, Myr,
             \end{array}
  \right.
  \eeq
  where $\tau_{age} $ stands for the age of the pulsar, provided the object is perfectly isolated and shielded from whatsoever external effects.
  In calculating the ratio at $\tau_{age}=0,$ the initial values $\alpha^{(n=0)}_c = 3.5 \times 10^{-10},$  $\Omega^{(n=0)} = 1400/s$ and $B^{(n=0)}=10^{13}~$Gauss have been used (see \cite{Hansel1999} for further details). The values at $\tau_{age}=1000, 10000~$yrs
  and $ \tau_{age}~ =~10\, Myr$  are chosen to enable partial comparison with observations of the glitch events of Vela and Crab pulsars.
  On the other hand the numerical values have been selected from the sequence $\{\alpha^{n}_c\}$ displayed in Fig.(8)  of \cite{HujeiratGlitch18}.\\
  For determining $R^n_{SB}$ we use the current glitch-observation of the Vela and Crab and try to extrapolate them to other epochs.
  Following the analysis and Eqs. (13, 14, 15) in \cite{HujeiratGlitch18}, the correlation of the inertia of the ambient dissipative media, $I_{AM},$
   of the Crab and Vela pulsars reads:
   \beq
                                 I^{Vela}_{AM} \approx 10^{-2} I^{Crab}_{AM},
   \eeq
  whereas the requirement that ejected rotational energy from the core of the Vela pulsar in the form of vortices should  be observationally noticeable,
  implies that  $R^{now, Vela}_{SB}\geq 10^{-2} R^{Vela}_\star.$ This implies that $\delta R^{now, Vela}_{SB}\approx \DD{5}{2} R^{now, Crab}_{SB}.$ Consequently,
  the cosmic values of $R^{n}_{SB}$ can be summarized as follows:
      \beq
        \hspace*{-2mm}
      \delta R^n_{SB}=
      \left\{
       \begin{array}{ll}
              \mathcal{O} (10^{-7}~cm)  & \tau_{age}=0 \\
              \mathcal{O} (10^{-4}~cm)  &  \tau_{age} =~ 1000~ yrs/Crab \\
               \mathcal{O} (10^{-2}~cm)  &  \tau_{age} =~ 10000~ yrs/Vela\\
               \mathcal{O} (1~ cm)  & \tau_{age}~ =~ 10\, Myr,
             \end{array}
  \right.
  \eeq
  An important question may be posed here:  \\
  \textbf{What is the nature of  force in the boundary layer that is capable of triggering glitches?}\\
  Here we argue that the strong nuclear force (SNF) is the deriving force that is capable of  changing instantly both
  the physical properties of the quantum fluid as well as the topology of spacetime in the BL.
  \begin{itemize}
  \item  Following \cite{HujeiratGlitch18}, the core mimics a super-baryon (SB), in which the enclosed quarks are
        shielded by a gluon-cloud. The SB interacts then  with overlaying supranuclear dense neutrons via vector mesons:
        the messengers of the SNF.  In turn, the SNF locks the neutrons gradually to the core and enforces them to rotate rigidly with the core.
\item  The SNF generally operates effectively on nucleon length scales, i.e. when $\delta R^n_{BL} \approx R^n_{SB} = \mathcal{O}(1)~fm.$
                   However, the inertia of a core of a neutron-size would be  approximately 100 orders of magnitude smaller than that of the ambient medium. In this case, its rotational and thermal decoupling from the ambient medium would be almost impossible.
                  On the other hand, in order for a SuSu-core to survive and affect the dynamics of  the ambient medium, its inertia must have a  minimum value.
                   Here we recall that the initial frequency of a newly born pulsar is 1400 /s  approximately \cite{Hansel1999}. In order for
                   a SuSu-core to be able to eject one single vortex, its radius must be larger than $10^3\,$cm, hence its minimum mass: $\mathcal{M}^{t=0}_{SB} \geq 2.58\times 10^{24}~$g.
                  { Indeed, the above tabulated values of $\delta R^n_{BL} $ show that $ \delta R^n_{BL}$ correlates nicely  with the size of the enclosed SuSu-core and it may even become of macroscopic size at the end of pulsars luminous lifetime.}
\item The essential question that rises is:\\
 \textbf{What keeps the core dynamically stable against the weight of the overlying massive shells, or equivalently  against  compression 
     by the surrounding strongly curved spacetime?}
   As the gravitational potential inside the core is constant, then the medium has zero stratification due to gravity.
   In this case the only opposing force is the pressure gradient at the interface between SuSu-cores and the ambient compressible dissipative medium.
   The pressure gradient here could be seen as the repulsion force due to Pauli exclusion principle.  In this case,  one possible
   configuration would be to have the gluon-quarks uniformly distributed on the surface of the SuSu-core, forming thereby
   a surface tension that is capable of confining the quarks whilst opposing further compressions.

        \item  As the neutron's density in the BL approaches the critical value, $\rho_{crit},$  and  as the chemical potential becoming closer
                    to the upper-limit, $\mu_0,$ the EOS should converge to $P \rightarrow a_0\, \bar{n}^2, $ where $\bar{n}$ is the number density.
                    This limiting EOS corresponds to purely incompressible fluids, implying therefore that the topology of the embedding spacetime
                    must  change into a flat spacetime.  In this case, two instantaneous  reactions in the BL are expected to occur (see Fig.\ref{3StageScenario}
                    a detailed description):
                    \begin{description}
                      \item[(a)] The BL  merges instantly with the SuSu-core to form a single quantum entity, thereby increasing its size and effective mass.
                      \item[(b)] The difference $\Delta \Omega = \Omega_n - \Omega_{n+1}$ between the rotational frequency of the combined BL-core system, i.e.  the newly formed entity,  and that of the ambient medium must have surpassed a critical quantum value.  In this case, the system  undergoes a transition into a lower energy state by  promptly expelling a certain number of vortices into the ambient medium.
                    \end{description}
  \end{itemize}
  \item  Based on the here-presented scenario, the mechanisms underlying the glitch phenomena in pulsars and young neutron stars are due to
            dramatic changes of the physical conditions of the matter as well as of the topology of spacetime in the BL.  Here the system operates on two different
             time scales: a relatively long times scale that correlates with size of the SuSu-core (e.g. approximately two year in the case of the Crab and Vela) and quantum events that occur instantly. The long-term changes can be listed as follows:
      \beq
        \hspace*{-2mm}
          \overbrace{\textrm{Slow processes}}^{\tau_p \approx ~2~yr}:
         \left\{
           \begin{array}{lll}
           EOS                & \rightarrow  & P = \varepsilon\\
            Compressible & \rightarrow  & Incompressible\\
            Dissipative     &\rightarrow   & Superfluid, \\
             \end{array}
         \right.
  \eeq
  while the fast actions are:
      \beq
        \hspace*{-2mm}
          \overbrace{\textrm{Fast actions}}^{\tau_p \leq 10^{-10} s}:
         \left\{
           \begin{array}{lll}
           Neutron~fluid      & \rightarrow & SuSu\\
          Curved~ST & \rightarrow & Flat~ST        \\
          E_{kin}               & \rightarrow & \downdownarrows E_{kin} \\
                                      & \Leftrightarrow & \downdownarrows N_{vortices}.
        \end{array}
         \right.
  \eeq


    \item   As the ambient compressible and dissipative medium cools and spins down and subsequently converts into
              SuSu matter, the core becomes more massive and larger. By the end of  luminous lifetime of the pulsar, the whole
              object should have entirely metamorphosed into a SuSu-object, thereby doubling its initial mass  and yielding
              radius that roughly coincides the event horizon. For remote observers the object becomes practically invisible and therefore
             indistinguishable from stellar black  holes.\\
             This evolutionary scenario may explain nicely, why neither black holes nor neutron stars have ever been observed in the mass range
             $ \{2.5~\mathcal{M}_\odot \leq \mathcal{M} \leq5~\mathcal{M}_\odot \}$ as well as why the NS-merger in GW170817 \cite{NSMergerLIGO} did not collapse  into a stellar BH.
\end{itemize}


{\bf{Acknowlegment}}
The calculations have been carried out using the computer cluster of the IWR, University of Heidelberg. RS is supported by KAUST baseline research funds.

 \end{document}